\documentclass{article}
\usepackage{spconfa4, amsmath, amsfonts, graphicx, bm, booktabs, multirow, xspace, xcolor}
\usepackage[sort]{cite}
\usepackage[hidelinks]{hyperref}
\usepackage{flushend}

\newcommand{\tsixty}{$\mathrm{T}_{60}$\xspace}
\newcommand{\cfifty}{$\mathrm{C_{50}}$\xspace}

\title{Blind Acoustic Parameter Estimation Through Task-Agnostic Embeddings Using Latent Approximations}
\name{Philipp G\"{o}tz$^{1}$, Cagdas Tuna$^{2}$, Andreas Brendel$^{2}$, Andreas Walther$^{2}$, and Emanu\"{e}l A. P. Habets$^{1}$\thanks{$^\dag$A joint institution of the Friedrich-Alexander-University Erlangen-N\"{u}rnberg (FAU) and Fraunhofer IIS.
\newline Corresponding author: philipp.goetz@audiolabs-erlangen.de}}
\address{$^1$International Audio Laboratories Erlangen$^\dag$, Germany.
\\ $^2$Fraunhofer Institute for Integrated Circuits IIS, Erlangen, Germany.}
\begin{document}
\ninept
\setlength{\textfloatsep}{12pt}
\maketitle
\begin{abstract}
We present a method for blind acoustic parameter estimation from single-channel reverberant speech. The method is structured into three stages. In the first stage, a variational auto-encoder is trained to extract latent representations of acoustic impulse responses represented as mel-spectrograms. In the second stage, a separate speech encoder is trained to estimate low-dimensional representations from short segments of reverberant speech. Finally, the pre-trained speech encoder is combined with a small regression model and evaluated on two parameter regression tasks. Experimentally, the proposed method is shown to outperform a fully end-to-end trained baseline model.
\end{abstract}
\section{Introduction}

Blind acoustic parameter estimation has been researched for over two decades \cite{eaton2016estimation}. Initially, statistical models of an observed reverberant signal, such as signal decay rate distributions \cite{ratnam2003blind,wen2008blind,eaton2013noise,parada2016single,talmon2013blind}, were used.
With the rise in popularity of deep learning methods, researchers have employed deep neural networks (DNN) and advanced the state-of-the-art by a significant margin. Gamper and Tashev proposed a convolutional neural network (CNN) that operates on Gammatone spectrograms of reverberant, noisy speech and estimates the reverberation time \tsixty \cite{gamper2018blind}. Subsequent works have investigated extensions towards different model architectures \cite{deng2020online}, different acoustic parameters \cite{srivastava2021blind,genovese2019blind,saini2023blind} and dynamic acoustic conditions \cite{gotz2023online}.

Most previous deep learning-based work focuses on estimating parameters from reverberant, noisy audio signals in an end-to-end manner. These task-specific models often fail to generalize to different parameters. Leveraging recent advances in self-supervision and representation learning, a number of works have aimed at learning general, task-agnostic representations from audio signals that are suitable in a range of downstream scenarios by solving various auxiliary tasks \cite{saeed2021contrastive,gong2022ssast}. In a previous contribution \cite{goetz_contrast}, we proposed a method for learning a task-agnostic representation of the acoustic environment through a supervised contrastive classification task combined with a novel sampling strategy. This method clusters reverberant speech samples with the same room impulse response (RIR) in latent space while separating those with different RIRs.
Additionally, we proposed a method for constructing these contrasts during training, which we demonstrated has a significant influence on the extracted embeddings. As a result, the learned representation becomes invariant to the anechoic source component, encapsulating only information about the acoustic channel. Although the representation remains (downstream) task-agnostic, the classification of different RIRs into distinct classes implies that the learned latent space does not reflect similarities between different RIRs. In this work, we propose an approach to learning a latent representation of a reverberant speech segment that is explicitly structured based on the underlying RIR. We demonstrate the advantage of the learned representations in two common downstream tasks.


We aim to estimate the acoustic parameters reverberation time \tsixty and clarity index \cfifty from monaural reverberant speech. Both parameters can be computed from the RIR, which generated the reverberant signal. However, due to the varying length and stochastic nature of the RIRs, their blind estimation from a single-channel signal represents a challenging problem. In this work, we follow a three-stage approach. First, we train a DNN to learn a compact, smooth, latent representation of the RIR, which then serves as a target for a second DNN that operates on reverberant speech. In the final stage, the approximated RIR latent representations are used as input to a third, shallow DNN that estimates the acoustic parameters of interest. This separation of the model into three consecutive stages offers considerable flexibility in the design of its individual components. In a recently proposed method for blind RIR estimation \cite{lee2023yet}, which is closely related to our approach, compact latent representations of RIRs are also learned in an initial step. However, the subsequent approximation of the RIR latents from reverberant speech is formulated as an auto-regressive token generation task, relying on vector-quantized RIR latents. Instead, we circumvent the requirement for vector quantization and its associated challenges by formulating the RIR latent approximation as a regression problem.

The remainder of this paper is organized as follows. In Sections~\ref{sec:problem} and \ref{sec:method}, the problem is stated, and the proposed method is described in detail. In Section~\ref{sec:data}, the data generation is outlined. In Sections~\ref{sec:experiments} and \ref{sec:discussion}, the experimental evaluation is presented and discussed. Section \ref{sec:conclusion} concludes this work.

\section{Problem Formulation}
\label{sec:problem}
We consider the observed reverberant discrete-time signal $y[n]$ to consist of the convolution of an anechoic source component $x[n]$ with RIR $h[n]$ of length $L$, and an additive, uncorrelated noise term $v[n]$:
\begin{equation}
    y[n]=\sum_{m=0}^{L-1} h[m] \, x[n-m] + v[n].
\end{equation}
In this study, both reverberant speech and RIRs are represented as log-magnitude mel spectrograms, denoted by $\bm{Y}$ and $\bm{H}$, which are matrices in $\mathbb{R}^{F\times T}$. Here, $F$ and $T$ represent the frequency and time dimensions, respectively.

We aim to develop a general approach to estimate acoustic parameters from the reverberant speech signal $\bm{Y}$ blindly, i.e., without access to the RIR. We demonstrate the effectiveness of the approach by jointly estimating the acoustic parameters \tsixty and \cfifty in seven-octave bands with center frequencies $f_c\in\left\{125, 250, 500, 1\mathrm{k}, 2\mathrm{k}, 4\mathrm{k}, 8\mathrm{k}\right\}$ Hz.

\section{Proposed Method}
\label{sec:method}
\begin{figure}
    \centering
    \includegraphics[width=\columnwidth]{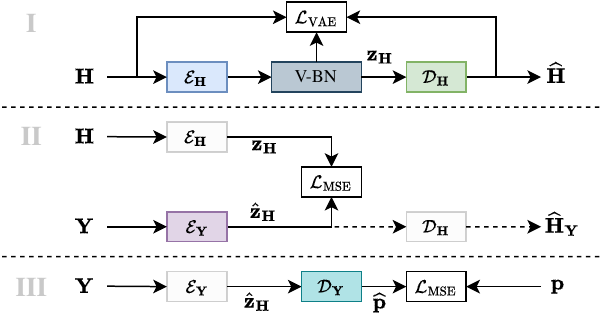}
    \caption{Overview of the proposed parameter estimation method: Stage I contains the RIR-VAE that learns compact latent representations in the variational bottleneck (V-BN), Stage II encompasses the approximation of the RIR latents from reverberant speech, in Stage III the approximated latent representations are used as input from which two acoustic parameters are estimated. The model components which are trained in each stage are highlighted.}
    \label{fig:model_overview}
\end{figure}

In the following, we describe the three different stages of the proposed parameter estimation method. An overview of the three stages of the method is shown in Fig.~\ref{fig:model_overview}.

\subsection{Stage I: RIR-VAE}
In the first stage, we learn compact latent representations of a set of RIRs $\mathcal{H}$. To this end, we train a variational auto-encoder \cite{kingma2013auto} consisting of encoder network $\mathcal{E}_\mathbf{H}$ and decoder network $\mathcal{D}_\mathbf{H}$ by maximizing the Evidence Lower Bound Objective (ELBO) with respect to their parameters $\bm{\phi}$ and $\bm{\theta}$:
\begin{multline}\mathcal{L}_{\mathrm{VAE}}=\\\mathbb{E}_{q_\phi\left(\bm{z_H}|\bm{H}\right)}\log p_\theta(\bm{H}|\bm{z_H}) - \mathrm{KL}\left\{ q_\phi(\bm{z_H}|\bm{H}) \, || \, p(\bm{z_H}) \right\},
\end{multline}
where $\mathrm{KL}$ denotes the Kullback-Leibler Divergence between the Gaussian prior $p(\bm{z})\sim\mathcal{N}(\bm{0}_D,\bm{I}_{D\times D})$ on the latent representation $\bm{z}$ and the surrogate posterior $q_\phi(\bm{z}|\bm{H})\sim\mathcal{N}(\bm{\mu},\bm{\Sigma})$, parameterized by a mean vector $\bm{\mu}\in\mathbb{R}^D$ and a covariance matrix $\bm{\Sigma}=\mathrm{diag}\left\{\bm{v}\right\}\in\mathbb{R}^{D\times D}$. The mean and variance parameters for the surrogate posterior $q_\phi(\bm{z}|\bm{H})$ are estimated by the encoder network $\mathcal{E}_\mathbf{H}$.

\begin{figure}
    \centering
    \includegraphics[width=\columnwidth]{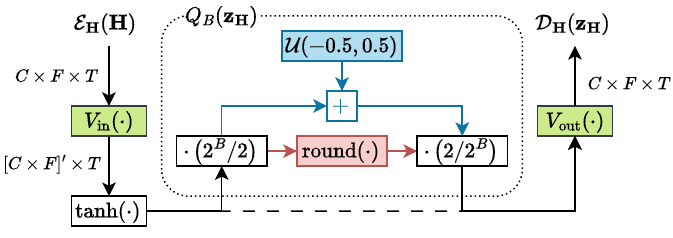}
    \caption{Architecture of the variational bottleneck used in Stage I, including projection layers in and out of the latent space and optional scalar quantization. Noise drawn from a uniform distribution is added during training (blue path), and scalar quantization using a rounding operation is applied during inference (red path).}
    \label{fig:bottleneck}
\end{figure}

We extend the architecture of the variational bottleneck by linear projection layers in and out of the latent space, $V_{\mathrm{in}}:\mathbb{R}^{CF}\rightarrow\mathbb{R}^{D}$ and $V_{\mathrm{out}}:\mathbb{R}^{D}\rightarrow\mathbb{R}^{CF}$, that operate along the combined channel and frequency dimensions \cite{yu2022vectorquantized}. Additionally, we add an optional quantization module $Q_B$ in which we add noise drawn from a uniform distribution to the latent features during training as a stand-in for the scalar quantization to $B$ bits, applied during inference \cite{ballé2018variational,mentzer2024finite} (cf. Fig. \ref{fig:bottleneck}).
The projection layers and scalar quantization efficiently constrain the representation capacity of the RIR-VAE and control the detail captured in $\mathbf{z}_\mathbf{H}$, the target for the next stage. In Section \ref{ssec:latent_comppression}, these two design choices are investigated with respect to the parameter estimation performance in an ablation study.

\subsection{Stage II: Speech Encoder}
In the second stage of the proposed method, we train an encoder $\mathcal{E}_\mathbf{Y}$ to extract embeddings from reverberant speech that approximate the latent representations of the RIRs that were used to reverberate the anechoic speech signal, i.e., $\mathcal{E}_\mathbf{Y}(\bm{Y})\approx\mathcal{E}_\mathbf{H}(\bm{H})$. This is achieved by minimizing the mean squared error (MSE) between speech and RIR representations in the latent space:
\begin{equation}
    \mathcal{L}_{\mathrm{MSE}}=\left\|\hat{\bm{z}}_{\bm{H}}-\bm{z_H}\right\|_2^2 \;\;\mbox{with}\;\; \hat{\bm{z}}_{\bm{H}}=\mathcal{E}_\mathbf{Y}(\bm{Y}).
\end{equation}

Inspired by the architectures used in \cite{luo2019conv,herzog2022ambisep}, we use a CNN encoder to extract local time-frequency features from the reverberant log mel spectrogram. The frequency and channel features are then concatenated and serve as input tokens to a series of transformer encoder layers that model long-term dependencies. Temporal aggregation and a linear projection into the RIR latent space yield the final approximated RIR latent representation.

\subsection{Stage III: Downstream Model}
In the last stage, we train a lightweight downstream model that uses the speech embeddings as input to estimate acoustic parameter vectors $\bm{p}$. The model is trained by minimizing the MSE between the ground-truth parameters and the estimated parameters, i.e.,
\begin{equation}
    \mathcal{L}_{\mathrm{MSE}}=\left\|\bm{\widehat{p}}-\bm{p}\right\|_2^2.
\end{equation}
The speech encoder is kept fixed during downstream training, i.e., no task-specific fine-tuning is carried out.

\section{Dataset}
\label{sec:data}
We created a dataset of $6,269$ measured RIRs from individual resources: the ACE challenge dataset \cite{eaton2016estimation}, the EchoThief database \cite{echothief}, the IKS Aachen Impulse Response database \cite{jeub2009binaural}, the OpenAir database \cite{murphy2010openair}, the WDR broadcast studio impulse response dataset \cite{lubeck2021high}, the spatial impulse response dataset gathered at the Technical University of Ilmenau \cite{klein2022dataset}, the RIR dataset from the MIT reverberation survey \cite{traer2016statistics} and the RIR dataset from the variable acoustic room at the Aalto Acoustics Labs \cite{https://doi.org/10.5281/zenodo.6985104}. The speech used in this study is sourced from a corpus of multi-lingual speech segments recorded in an anechoic chamber \cite{speech_dataset}.

Before creating reverberant speech signals, we divided all RIRs and speech signals into two separate subsets: one for Stages I and II and another for Stage III of the proposed method. This split guarantees that downstream testing is performed exclusively on speech and RIRs that have not been previously seen by the model. Each of the two splits is then again separated into disjoint subsets for training, validation, and testing. For Stages II and III, we generated approximately 20 hours of reverberant speech in 4-second segments each.

All signals in this study, RIRs and speech, are processed at a sampling rate of $16\,\mathrm{kHz}$ as magnitude spectrograms with a mel frequency scale. Specifically, we use a short-time Fourier transform with a window length of $64$ samples and hop sizes of $16$ samples for RIRs and $32$ samples for reverberant speech, respectively. The linear spaced frequency bins are mapped in each time frame to $16$ mel sub-bands. The reverberant speech is transformed to log-mel spectrograms and scaled such that each sample has zero-mean and unit standard deviation.

The ground truth for \tsixty is obtained using the single-slope variant of the DecayFitNet \cite{gotz2022neural}, and the ground truth for \cfifty is directly computed from the band-limited RIRs $h_b[n]$ according to:
\begin{equation}
    \mathrm{C_{50}^{(b)}} = 10 \log_{10}\left(\frac{\sum_{n=0}^{M-1}h_b^2[n]}{\sum_{n=M}^{L-1}h_b^2[n]}\right),
\end{equation}
where $L$ denotes the length of the RIR, $n=0$ denotes the time index of arrival of the direct sound, and $M$ corresponds to the time index $50\,\mathrm{ms}$ after the direct sound.

\section{Experiments}
\label{sec:experiments}
\begin{figure}[ht]
    \centering
    \includegraphics[width=\columnwidth]{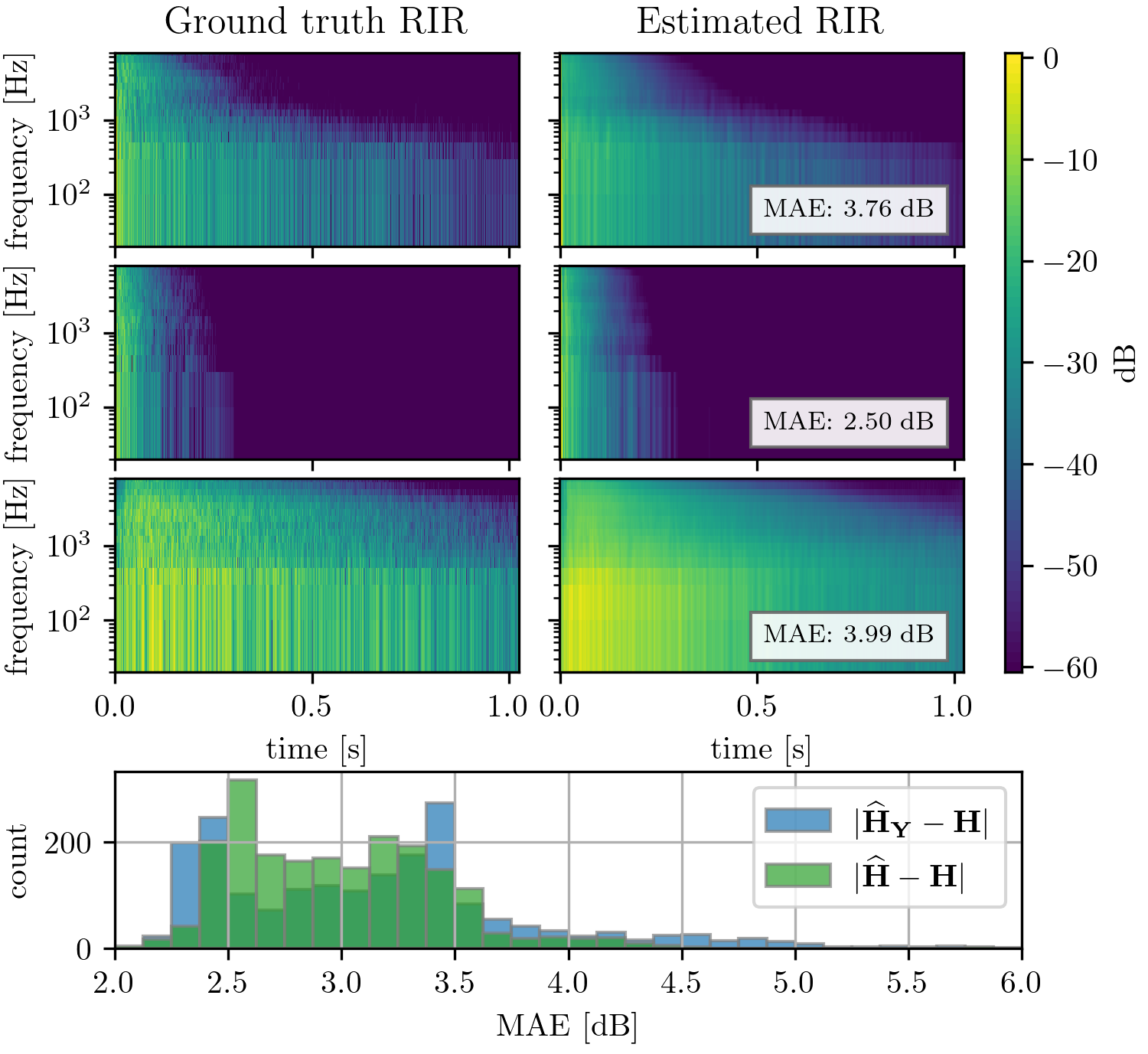}
    \caption{Three examples of mel spectrograms of RIRs, decoded from approximated RIR latents. The left column shows the ground truth RIR, the right column shows the estimate from speech and the corresponding MAE. Below the examples are histrograms of the RIR reconstruction error, shown for the true and estimated RIR latents.}
    \label{fig:rir-examples}
\end{figure}

We evaluated the proposed method in terms of performance metrics across two downstream tasks and the reconstruction quality of decoded RIR latent approximations. We compared the proposed method for both acoustic parameters to a fully end-to-end trained baseline and to a previously proposed method for learning task-agnostic representations \cite{goetz_contrast}. The same dataset was used to train the models used for the proposed, contrastive, and end-to-end approaches. Before evaluating downstream task performance, we highlight the proposed method's ability to obtain RIR estimates in the mel-frequency domain from approximated RIR latents. Figure \ref{fig:rir-examples} shows three examples of RIRs that were estimated from speech based on unquantized RIR latents. Some fine temporal structure in the mel spectrogram is lost through the two model stages, but the overall characteristics of the RIR and even some of the individual details are preserved. Below the three examples, a histogram of the reconstruction errors in the test dataset indicates a bimodal distribution. We note that a large portion of the errors are distributed over a fairly limited range and conjecture that the lower limit at approximately $2.2\,\mathrm{dB}$ may be a result of the RIR latent compression in the variational bottleneck. However, this hypothesis would require further empirical testing to validate.

\subsection{Downstream Task Performance} 
Table \ref{tab:results} shows the mean absolute error (MAE), Pearson correlation coefficient (PCC), and estimation bias (BIAS) -- metrics commonly used for evaluation \cite{eaton2016estimation} -- for both acoustic parameters according to the proposed method, the contrastive approach, the end-to-end baseline and for the non-blind case where the downstream model was trained on true RIR latents obtained using the RIR-VAE $\mathcal{E}_{\bm{H}}$. All results presented in Table \ref{tab:results} were obtained from the average of eight-fold cross-validation runs using unquantized RIR latents with the largest dimension considered, i.e., $D=512$. In the case of \tsixty, the proposed method generally shows the lowest MAE across all frequency bands, indicating more accurate estimations compared to the contrastive learning approach and the end-to-end baseline. Pearson correlation coefficients can be considered high for all three approaches, particularly for the contrastive and proposed methods, which exhibit a clear advantage over the end-to-end baseline in most octave bands. Finally, the estimation bias can be considered small for the three blind methods, with the end-to-end baseline exhibiting the smallest offset. The proposed method has a slight tendency to underestimate \tsixty while the contrastive training tends to slightly overestimate. For \cfifty, the proposed method again shows lower MAE in most frequency bands, with a notable improvement compared to the other approaches, especially in the higher octaves. PCC values are very high for the proposed method across all bands, typically surpassing those of the contrastive approach and matching or slightly exceeding the end-to-end baseline. The observed bias indicates a slight overestimation of \cfifty consistent across all bands for the proposed method, whereas the contrastive learning approach exhibits the smallest bias. Compared with the end-to-end baseline, which results in the largest bias, the bias of the proposed method can be deemed acceptable.

\begin{table*}[t]
    \centering
    {\renewcommand{\arraystretch}{1.0}
    \resizebox{\textwidth}{!}{
    \begin{tabular}{p{0.25cm} c c c c c c c c c c c c c c c}
    \toprule
         & & \multicolumn{4}{c}{MAE} && \multicolumn{4}{c}{PCC} && \multicolumn{4}{c}{BIAS}  \\
         \cmidrule{3-6} \cmidrule{8-11} \cmidrule{13-16}
         & $f_c\,[\mathrm{Hz}]$ & Proposed & Contr. & E2E & UB && Proposed & Contr. & E2E & UB && Proposed & Contr. & E2E & UB \\
         \midrule

\multirow{7}{*}{\tsixty} & $125$ & $\bm{0.130}$	& $0.184$ & $0.151$ & $\color{gray}{0.073}$ && $\bm{0.911}$ & $0.818$ & $0.854$ & $\color{gray}{0.969}$ && $\bm{0.005}$ & $-0.029$ & $0.037$ & $\color{gray}{-0.006}$\\
& $250$ & $\bm{0.107}$ & $0.153$ & $0.109$ & $\color{gray}{0.058}$ && $\bm{0.940}$ & $0.867$ & $0.886$ & $\color{gray}{0.982}$ && $0.017$ & $-0.021$ & $\bm{0.001}$ & $\color{gray}{-0.006}$ \\
& $500$ & $\bm{0.101}$ & $0.169$ & $0.113$ & $\color{gray}{0.054}$ && $\bm{0.956}$ & $0.865$ & $0.936$ & $\color{gray}{0.987}$ && $0.011$ & $-0.018$ & $\bm{0.003}$ & $\color{gray}{-0.003}$ \\
& $1\,\mathrm{k}$  & $\bm{0.095}$ & $0.186$ & $0.119$ & $\color{gray}{0.050}$ && $\bm{0.962}$ & $0.853$ & $0.947$ & $\color{gray}{0.990}$ && $0.011$ & $-0.014$ & $\bm{0.002}$ & $\color{gray}{-0.003}$ \\
& $2\,\mathrm{k}$  & $\bm{0.091}$ & $0.182$ & $0.111$ & $\color{gray}{0.051}$ && $\bm{0.963}$ & $0.845$ & $0.951$ & $\color{gray}{0.989}$ && $0.007$ & $-0.011$ & $\bm{-0.004}$ & $\color{gray}{-0.001}$ \\
& $4\,\mathrm{k}$  & $\bm{0.085}$ & $0.158$ & $0.104$ & $\color{gray}{0.051}$ && $\bm{0.952}$ & $0.830$ & $0.928$ & $\color{gray}{0.982}$ && $\bm{0.008}$ & $-0.006$ & $-0.010$ & $\color{gray}{-0.003}$ \\
& $8\,\mathrm{k}$  & $\bm{0.089}$ & $0.132$ & $0.109$ & $\color{gray}{0.052}$ && $\bm{0.901}$ & $0.774$ & $0.849$ & $\color{gray}{0.966}$ && $0.008$ & $-0.003$ & $\bm{0.000}$ & $\color{gray}{-0.002}$\\
\midrule
\multirow{7}{*}{\cfifty} & $125$ & $\bm{2.99}$ & $4.22$ & $3.34$ & $\color{gray}{2.33}$ && $\bm{0.913}$ & $0.817$ & $0.899$ & $\color{gray}{0.946}$ && $-0.25$ & $-0.12$ & $\bm{0.05}$ & $\color{gray}{-0.10}$ \\
& $250$ & $2.34$ & $3.88$ & $\bm{2.28}$ & $\color{gray}{1.82}$ && $0.949$ & $0.844$ & $\bm{0.955}$ & $\color{gray}{0.966}$ && $-0.36$ &  $\bm{0.14}$  & $0.24$ & $\color{gray}{-0.01}$ \\
& $500$ & $\bm{2.09}$ & $3.71$ & $2.34$ & $\color{gray}{1.82}$ && $\bm{0.952}$ & $0.826$ & $0.941$ & $\color{gray}{0.957}$ && $-0.37$ & $\bm{0.05}$  & $0.31$ & $\color{gray}{-0.08}$ \\
& $1\,\mathrm{k}$ & $2.05$ & $3.25$ & $\bm{1.98}$ & $\color{gray}{1.73}$ && $0.953$ & $\bm{0.960}$ & $0.931$ & $\color{gray}{0.961}$ && $-0.20$ & $\bm{0.04}$ & $0.46$ & $\color{gray}{-0.07}$ \\
& $2\,\mathrm{k}$ & $\bm{2.08}$ & $3.70$ & $2.16$ & $\color{gray}{1.90}$ && $0.955$ & $\bm{0.962}$ & $0.903$ & $\color{gray}{0.960}$ && $-0.16$ & $\bm{-0.07}$ & $0.51$ & $\color{gray}{-0.05}$ \\
& $4\,\mathrm{k}$ & $\bm{2.23}$ & $3.84$ & $2.40$ & $\color{gray}{2.06}$ && $0.951$ & $\bm{0.956}$ & $0.898$ & $\color{gray}{0.957}$ && $-0.35$ & $\bm{-0.18}$ & $0.45$ & $\color{gray}{-0.12}$ \\
& $8\,\mathrm{k}$ & $\bm{2.72}$ & $4.43$ & $2.78$ & $\color{gray}{2.60}$ && $0.933$ & $\bm{0.947}$ & $0.893$ & $\color{gray}{0.939}$ && $-0.15$ & $\bm{0.05}$ & $0.72$ & $\color{gray}{-0.20}$ \\
\bottomrule
    \end{tabular}}}
    \caption{A comparison of downstream performance metrics across the three considered methods: the proposed multi-stage approach (Proposed), the contrastive pre-training (Contr.)\cite{goetz_contrast}, the end-to-end (E2E) baseline, and the non-blind upper bound $\mathcal{D}_{\bm{H}}(\bm{z_H})$, denoted (UB). The best results for each metric within each octave band are highlighted in bold.}
    \label{tab:results}
\end{table*}

In Table \ref{tab:results}, the upper bound (UB) results, represented in grey, serve as an experimental benchmark as they were obtained by training the downstream model $\mathcal{D}_{\bm{Y}}$ on true RIR latents $\bm{z_H}$. Across both acoustic parameters, \tsixty and \cfifty, and nearly all frequency bands, the proposed method's results are not markedly far from this non-blind upper bound, which is indicative of the richness of the latent representations that are approximated from reverberant speech signals. 

\subsection{RIR Latent Compression and Quantization}
\label{ssec:latent_comppression}
\begin{figure}[!t]
    \centering
    \includegraphics[width=\columnwidth]{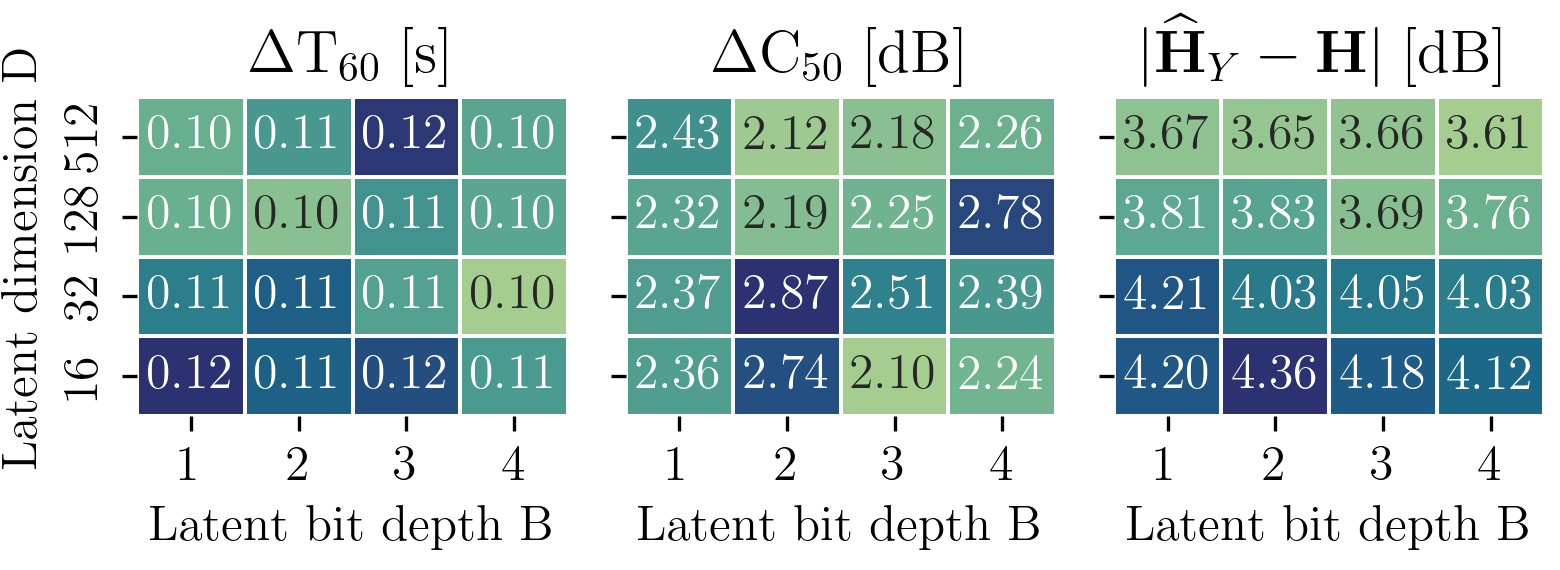}
    \caption{Ablation results of the latent compression in the RIR-VAE bottleneck. Left and center: frequency-averaged \tsixty and \cfifty estimation errors, right: mean-absolute error of RIR reconstruction.}
    \label{fig:latent-ablation}
\end{figure}

We studied the compression and quantization of RIR latents in the bottleneck of the RIR-VAE with respect to the resulting downstream performance and the RIR reconstruction quality from speech. We trained the RIR-VAE, the speech encoder, and the downstream model for $16$ different combinations of latent dimensionality ($D$) and quantization bit depth ($B$). Figure \ref{fig:latent-ablation} indicates that estimation performance for both acoustic parameters does not strongly depend on the configuration of the variational bottleneck. While \tsixty estimation is largely unaffected by the latent structure, a more noticeable degradation can be observed for \cfifty when channel and frequency dimensions are projected to a lower-dimensional subspace. 

However, the reconstruction of RIRs depends on both bottleneck parameters, with the latent dimensionality having the bigger impact. These results suggest that latent space quantization may be considered for acoustic parameter estimation in applications with memory constraints.

\vspace{-1em}
\section{Discussion}
\label{sec:discussion}
The proposed three-stage approach offers flexibility in designing each stage, including data representation, model architecture, and training process. However, the sequentiality of the approach also implies that the approximation of the RIR latents from speech strongly depends on the information they contain, meaning that each stage in the model can be considered as a bottleneck for its following stage.

\section{Conclusion}
\label{sec:conclusion}
We have presented a multi-stage framework for blind acoustic parameter estimation. Our method integrates a variational auto-encoder for learning compact representations of (transformed) RIRs and an encoder of reverberant speech signals that approximates RIR latent vectors. We demonstrate the richness of the learned representations across two common downstream tasks and investigate the flexibility of the approach through an ablation study. The experimental results presented in this work validate the efficacy of using approximated RIR latent representations for precise and reliable parameter regression, setting a promising direction for future research in blind acoustic parameter estimation.

\bibliographystyle{IEEEbib}
\bibliography{refs}

\end{document}